\documentclass[twocolumn,prd,showpacs]{revtex4}
\usepackage{graphicx}

\begin{document}

\newcommand{\re}{\mathop{\mathrm{Re}}}

\newcommand{\ud}{\mathrm{d}}

\renewcommand{\sp}{\Omega_{sp0}}
\newcommand{\ph}{\Omega_{ph0}}
\renewcommand{\L}{\Omega_{\Lambda 0}}
\newcommand{\w}{\Omega_{w0}}
\newcommand{\K}{\Omega_{K'0}}
\newcommand{\m}{\Omega_{m0}}
\newcommand{\rr}{\Omega_{r0}}
\newcommand{\st}{\Omega_{st0}}

\newcommand{\be}{\begin{equation}}
\newcommand{\ee}{\end{equation}}
\newcommand{\bea}{\begin{eqnarray}}
\newcommand{\eea}{\end{eqnarray}}

\title{Phantom cosmologies}

\author{Mariusz P. D\c{a}browski}
\email{mpdabfz@uoo.univ.szczecin.pl}
\affiliation{\it Institute of Physics, University of Szczecin, Wielkopolska 15,
          70-451 Szczecin, Poland}
\author{Tomasz Stachowiak}
\email{toms@oa.uj.edu.pl}
\affiliation{\it Astronomical Observatory, Jagiellonian University, 30-244
Krak\'ow, ul. Orla 171, Poland}
\author{Marek Szyd{\l }owski}
\email{uoszydlo@cyf-kr.edu.pl}
\affiliation{\it Astronomical Observatory, Jagiellonian University, 30-244
Krak\'ow, ul. Orla 171, Poland}

\date{\today}

\input epsf

\begin{abstract}

We discuss a class of phantom ($p < -\varrho$) cosmological models. Except
for phantom we admit various forms of standard types
of matter and discuss the problem of singularities for 
these cosmologies.
The singularities are different from these of standard matter cosmology 
since they appear for infinite values of the scale factor. 
We also find an interesting relation between the phantom models and
standard matter models which is like the duality symmetry
of string cosmology. 
 
\end{abstract}

\maketitle

\section{Introduction}

The observations of distant supernovae \cite{Perlmutter99,Riess98}
has given the evidence for accelerating universe filled with
matter which violates the strong energy condition $\varrho + 3p \geq 0$,
where $p$ is the pressure and $\varrho$ is the energy density.
However, more detailed check of data suggests that also the matter
which violates the weak energy condition $\varrho + p \geq 0, \varrho \geq 0$
is admissible at high confidence level
\cite{caldwell99,Hann02,Frampt03,FramTaka02,crooks03,McInnes,trodden1,trodden2}.

The physical background for strongly negative pressure matter
(phantom) may be looked for in string theory \cite{Frampt02}. The
point is that the group velocity of a wave packet is negative
which leads to the behaviour of the packet which is somewhat
unusual - namely, it moves in the direction opposite to the
momentum. In such circumstances a wave packet which leaves the
comoving volume transfers momentum into the volume which makes
negative contribution to the pressure \cite{Bastero01,Bastero02}.

Formally, one can get phantom by switching the sign of kinetic
energy of the standard scalar field Lagrangian, i.e., by taking
${\cal L} = - (1/2) \partial_{\mu} \phi \partial^{\mu} \phi -V(\phi)$
which gives the energy density $\varrho = - (1/2) \dot{\phi}^2 + V(\phi)$
and the pressure $p = -(1/2) \dot{\phi}^2 - V(\phi)$ and leads to $\varrho + p =
- \dot{\phi}^2 < 0$ \cite{caldwell99,nojiri031,nojiri032,nojiri033,Erickson,BIphantom,LiHao03,singh03}.
Phantom type of matter may also arise from a
bulk viscous stress due to particle production \cite{john88} or in
higher-order theories of gravity \cite{Pollock88}, Brans-Dicke and non-minimally 
coupled scalar field theories \cite{diego02}.

The cosmological models which allow for phantom matter appear 
naturally in the mirage cosmology of the braneworld scenario \cite{kiritsis} 
and in kinematically-driven quintessence (k-essence) models \cite{chiba}.

It has been shown that the matter which violates the strong energy
condition allows the cyclic (oscillating, non-singular) universes
\cite{Dabrowski96}. It seems that similar solutions should also appear 
for the phantom matter which violates the weak energy condition.

A different idea of a cyclic universe in which the contracting big crunch phase can be connected
with an expanding big bang phase \cite{bd95} has been revived recently in the
context of M-theory cosmology motivated ekpyrotic scenario
\cite{ekpyrotic}. In this reference 
it has been shown in Ref. \cite{khoury01} that
the only way to make a transition from a contracting phase to an
expanding phase in a flat universe is to violate the weak energy
condition and this gives another motivation for studying
phantom cosmologies.

Some other proposals which may have something in common with
phantom are Cardassian models in which one adds an extra power in
the energy density term to the Friedmann equation
\cite{freese02,freese03,zhu}, or modified Einstein-Hilbert action
models in which an extra term of arbitrary (both positive and negative) 
power of scalar curvature appears in the
gravitational action \cite{EHRminusn,Starobinsky,nojiri034}. Both these possibilities 
may be directly motivated by string/M-theory \cite{braneIa,Sahni}.

\section{Basic system of equations with phantom}

Phantom is a new type of cosmological fluid which has a very
strong negative pressure which violates the weak energy condition
\cite{Hann02,Frampt02,Frampt03}, i.e., it obeys the equation of state
\be
p = \left({\gamma - 1}\right) \varrho = w \varrho,
\ee
with negative barotropic index
\be
\gamma = w + 1 < 0 ,
\ee
and this implies that $p < - \varrho$.

In order to study phantom cosmologies, we start our discussion with the basic
system of equations for isotropic and homogeneous Friedmann universe which reads as
\bea
\label{perho1}
\kappa^2 \varrho & = & - \Lambda + 3 \frac{k}{a^2} + 3
\frac{\dot{a}^2}{a^2} ,\\
\label{perho2}
\kappa^2 p & = & \Lambda - 2 \frac{\ddot{a}}{a} - \frac{k}{a^2} -
\frac{\dot{a}^2}{a^2}  ,
\eea
where $a(t)$ is the scale factor, $k = 0, \pm 1$ the curvature
index, $\Lambda$ the cosmological constant, $\kappa^2 = 8\pi G$ is the Einstein constant.

Using equations (\ref{perho1})-(\ref{perho2}) one gets the Friedmann equation in the
form
\begin{equation}
\label{Friedroro}
\frac{\dot{a}^2}{a^2} = \frac{\kappa^{2}}{3} \rho - \frac{k}{a^{2}} +
\frac{\Lambda}{3}.
\end{equation}
After imposing conservation law $\varrho a^{3\gamma} = (3/\kappa^2) C_{\gamma}$ =
const., one gets the Eq. (\ref{Friedroro}) as follows
\begin{equation}
\label{FriedCCC}
\frac{1}{a^2} \left( \frac{da}{dt} \right)^2 =
\frac{C_{\gamma}}{a^{3\gamma}}  -
\frac{k}{a^2} + \frac{\Lambda}{3} .
\end{equation}
The cases which involve all types of cosmological fluids, i.e., with $\gamma = 4/3$ (radiation),
$\gamma = 1$ (dust), $\gamma = 2/3$ (cosmic strings), $\gamma = 1/3$ (domain walls) and $\gamma = 0$
(cosmological constant) have been exactly integrated in terms of elliptic functions
\cite{Dabrowski96}. In fact, a large class of oscillating (non-singular) solutions have
been found.

Now we extend this discussion of the exact cosmological solutions into the case of
phantom matter with $\gamma = -1/3$ (we will call it phantom) and $\gamma = -2/3$ (we will
call it superphantom). One should emphasize that the ``border'' between ``standard''
and phantom models is given by the value of barotropic index
\be
\gamma = 0 \hspace{0.5cm} (w = -1).
\ee
Due to the appearance of the higher than four powers of the scale
factor in the Friedmann equation one cannot integrate a general
case which involves phantom matter by elliptic functions.
However, a lot of interesting special solutions can be found
without using the elliptic functions and this is the matter of the present paper.
We also include stiff-fluid matter with $\gamma = 2$.

From the conservation law we have
\be
\label{gammag0}
\varrho \propto a^{-3\gamma} \hspace{0.5cm} {\rm for} \hspace{0.5cm} \gamma > 0
\ee
and
\be
\label{gammal0}
\varrho \propto a^{3\mid \gamma \mid} \hspace{0.5cm} {\rm for} \hspace{0.5cm} \gamma < 0
\hspace{0.5cm} {\rm (phantom)} .
\ee
From (\ref{gammal0}) it is clear that big-bang and
big-crunch singularities appear at infinite values of the scale factor for phantom
cosmologies. This is also obvious after studying the simplest solution of
(\ref{FriedCCC}) for $k = \Lambda = 0$, which reads
\be
a(t) \propto t^{\frac{2}{3\gamma}} ,
\ee
so that
\be
\varrho \propto t^{-2} .
\ee
In other words, taking $\gamma = -1/3$ (phantom) one has $a(t) \to \infty$
and $\varrho \to \infty$ if $t \to 0$, while $a(t) \to 0$ and $\varrho \to 0$ if
$t \to \infty$. On the other hand, in a standard case $\gamma = 1/3$,
for example, one has $a(t) \to 0$ and $\varrho \to \infty$ if $t \to 0$,
while $a(t) \to \infty$ and $\varrho \to 0$ if $t \to
\infty$.
Similarly, one can show that the curvature invariants are
proportional to the energy density and so they are divergent wherever
the density/scale factor diverges
\bea
R &=& \kappa^2 \left(3 \gamma - 2\right) \varrho; \nonumber \\
R_{\mu\nu}R^{\mu\nu} &=& \frac{\kappa^4}{4} \left[ \left(3\gamma -
2 \right)^2 + \frac{1}{3} \left( 3\gamma + 2 \right)^2 \right]
\varrho^2 .
\eea

Another interesting remark can be extracted from the equations (\ref{perho1})-(\ref{perho2}) and 
(\ref{gammag0})-(\ref{gammal0}) if we admit shear anisotropy $\sigma_0^2/a^6$ ($\sigma_0=$ const.) 
and consider non-isotropic Bianchi type IX 
models. Namely, for $\gamma < 0$, the shear anisotropy cannot dominate over the phantom matter
on the approach to a singularity when $a \to \infty$, i.e., we have 
\be
\varrho a^{3 \mid \gamma  \mid} > \frac{\sigma_0^2}{a^6} \hspace{0.5cm} {\rm for} \hspace{0.5cm} a \to \infty 
\ee
and this prevents the appearance of chaotic behaviour of the phantom cosmologies of the 
Bianchi type IX \cite{chaos1,chaos2}.

\section{Phantom models duality}
        
The system of equations (\ref{perho1})-(\ref{perho2}) can be presented in the form
of the nonlinear oscillator
\be
\label{oscillator}
\ddot{X} - \frac{D^2}{3} \Lambda X + D(D -1) k X^{1- 2/D} =
0 ,
\ee
after introducing the variables \cite{szydlo84}
\be
X = a^{D(w)}, \hspace{0.5cm} D(w) = \frac{3}{2} (1 + w) .
\ee
For flat $k=0$ models the oscillator (mathematical pendulum) is in the lower equilibrium
position provided $\Lambda < 0$ and in the upper equilibrium
position provided $\Lambda > 0$ 
It is also easy to notice that the equation (\ref{oscillator}) preserves its form under the
change
\be
D \rightarrow - D ,
\ee
or
\be
\gamma \to - \gamma \hspace{0.5cm} \left(  w \rightarrow -(w+2) \right) .
\ee

It appears that there is a duality (similar to the scale-factor
duality in pre-big-bang models which is motivated by superstring theory duality symmetries 
\cite{meissner91,superjim} ) between the scale
factor and its inverse $a \to a^{-1}$ for standard matter ($\gamma > 0$) and
phantom matter -- $\gamma < 0$ models. Namely, the standard 
models for $w = \gamma - 1$ are dual to  phantom models for
$w' = - (2 + w) = - \gamma - 1$ with respect to $\gamma = 0$ line. For
example the domain wall models $\gamma_w = 1/3$ are dual to phantom
models $\gamma = -1/3$ and cosmic string models ($\gamma_{cs} = 2/3$) are dual to superphantom models.
However, this simple duality is valid only for flat
models (see also the Refs. \cite{lazkoz1,lazkoz2} where the similar results for both flat and non-flat models 
have been obtained and in the non-flat case the analogy might perhaps be the non-abelian duality 
\cite{non-abelian}).
In the former case we have (for walls, phantom and $\Lambda < 0$)
\bea
a_w & = & \left(\sin{\frac{\mid D \mid^{\frac{1}{2}}}{\sqrt{3}} \mid \Lambda \mid t} \right)^{\frac{1}{2}} ,\\
a_{ph} & = & \left(\sin{\frac{\mid D \mid^{\frac{1}{2}}}{\sqrt{3}} \mid \Lambda \mid t} \right)^{-\frac{1}{2}} ,
\eea
so that we have
\be
a_w = a_{ph}^{-1}, \hspace{0.5cm} D_w = 1/2 = - D_{ph} .
\ee   
From these we can conclude that standard fluids like: dust $\gamma = 1$, radiation $\gamma = 4/3$, stiff-fluid
$\gamma = 2$ are dual to phantoms with $\gamma = -1$ ($p = -2 \varrho$), $\gamma = 4/3$, $\gamma = -2$ ($p = -3 \varrho$),
respectively.
  
It is important to notice that duality in scale factor does not lead to the avoidance of singularity in the energy density which
results from (\ref{gammag0})-(\ref{gammal0}) and shows that whatever behaviour of the scale factor the density diverges leading to
a big-bang singularity.

\section{Phantom cosmological models}

For the sake of a possible comparison with observational data we
introduce dimensionless density parameters \cite{AJIII,braneIa}
\bea
\label{Omegadef}
\Omega_{x0}  &=&  \frac{\kappa^2}{3H_0^2} \varrho_{x0} ,\\
\Omega_{K0}  &=&  \frac{K}{H_0^2a_0^2} ,\\
\Omega_{\Lambda_0}  &=&  \frac{\Lambda_0}{3H_0^2} ,
\eea
where $x \equiv sp, ph, w, cs, m, r, st$ (superphantom, phantom, domain walls, cosmic strings,
dust, radiation and stiff-fluid, respectively), $H = \dot{a}/a$, and
$q  =  - \ddot{a}a/\dot{a}^2$, while \cite{AJIII}
\be
\label{kaprime}
\Omega_{K'0} = \Omega_{cs0} - \Omega_{K0} ,
\ee
which means that the Friedmann equation (\ref{FriedCCC}) can be written down
in the form
\begin{equation}
\label{Om=1}
\Omega_{sp0} + \Omega_{ph0} + \Omega_{\Lambda_0} + \Omega_{w0} + \Omega_{K'0}
+ \Omega_{m0} + \Omega_{r0} + \Omega_{st0} = 1  .
\end{equation}

Now using the variables
\be 
\label{param}
y = \frac{a}{a_0}, \hspace{0.5cm} u = H_0 t ,
\ee
one turns the Eq. (\ref{FriedCCC}) into the form
\bea
\label{FRWbasic}
\left( \frac{dy}{du} \right)^2 &=& \Omega_{sp0}y^4 + \Omega_{ph0}y^3 + \Omega_{\Lambda_0}y^2
+ \Omega_{w0} y + \Omega_{K'0} \nonumber \\
&+& \Omega_{m0} y^{-1} + \Omega_{r0} y^{-2} +
\Omega_{st0} y^{-4} \equiv Q(y).
\eea

\subsection{Negative pressure fluids only phantom models}

General solutions of Eq.(\ref{FRWbasic}) are given in terms
of elliptic or hyperelliptic functions. Here we concentrate only
on exact elementary solutions leaving a general discussion for a
separate paper \cite{elliptic}. Firstly, we assume non-vanishing negative pressure
fluids only (except negative $\Lambda$-term which gives positive pressure), 
i.e., $\m = \rr = \st = 0$, so that we have
\begin{equation}
\label{negpres}
\left( \frac{\ud y}{\ud u} \right)^2 = \sp y^4 + \ph y^3 + \L y^2 + \w y + \K .
\end{equation}
Note that after a change of variables 
\be
p \equiv \frac{1}{y} ,
\ee
we have
\be
\left( \frac{\ud p}{\ud u} \right)^2 = \sp+\ph p + \L p^2 + \w p^3 +\K p^4 ,
\ee
which may also be useful for exact integration.

Now we consider some special cases and refer to their mathematical
discussion presented in the Appendix A.

\begin{figure}[h]
\includegraphics[angle=0,scale=.46]{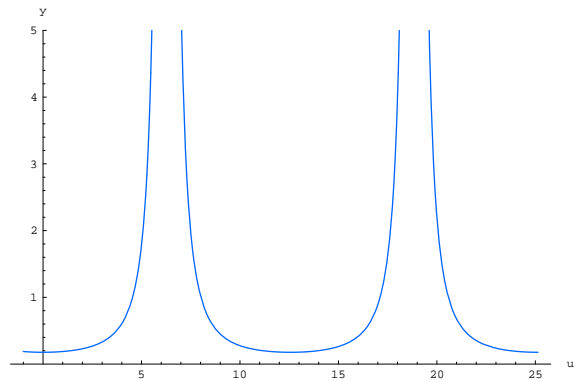}
\caption{The ``bounce'' solution (\ref{bounce}). The big-bang/big-crunch singularities
appear for $y \to \infty$ since $\varrho \propto y \to \infty$ at these points so that
this ``bounce'' in scale factor does not lead to singularity avoidance.}  
\label{fig02}
\end{figure}

\subsubsection{No-wall models with $\w=\K=0$}

This case falls under a general classification given in the Appendix A
with $a_2=\L\in{\bf R}$, $a_1=\ph\geq 0$,
and $a_0=\sp\geq 0$ (we exclude the possibility of both phantoms lacking, i.e.,
$\sp=\ph=0$).
The possible solutions are thus (cf. Appendix A):
\begin{description}
\item 1. $\sp=0,\;\ph>1$, and $\L<0$ (case 1.3.2 - cf. Fig. \ref{fig131}).
\item 2. $\sp>0,\;\ph>1-\sp$, and $\L<0$ (case 1.3.3).
\item 3. $\sp>0,\;\ph\in\left(0;2\sqrt{\sp}(1-\sqrt{\sp})\right)$,
and $\L>0$ (case 2.1).
\item 4. $\sp<1$, and $\L>0$ (case 2.2.3).
\item 5. $\sp=0,\;0<\ph<1$, and $\L>0$ (case 2.3.4).
\item 6. $\sp>0,\;0<\ph<1-\sp$, and $\L>0$ (case 2.3.5).
\end{description}

Most of these solutions (cases 3, 4, 5, 6) pass through zero in $x$,
so there arise infinite values of $y$ for finite times. Bearing in mind
Eq. (\ref{gammal0}) one can notice that this corresponds the energy density
approaching zero. However, in the cases 1 and 2
one obtains cyclic ``bounce'' in scale factor $y$ solution (which is also a ``bounce'' 
in the energy density $\varrho$ but not singularity avoidance since $\varrho \to \infty$ 
at finite $u$) of the form:
\begin{equation}
\label{bounce}
        y=\frac{2\L}{\sqrt{\ph^2-4\sp\L}\sin[\sqrt{|\L|}(u-u_0)]-\ph}
\end{equation}
which is shown in Fig. \ref{fig02}. In fact, there is a
competition between the positive pressure of a negative cosmological
term and the negative pressure of a phantom, but one is not able to avoid singularity
on the same basis as with negative cosmological constant and domain walls \cite{Dabrowski96}.

\begin{figure}[h]
\includegraphics[angle=0,scale=.46]{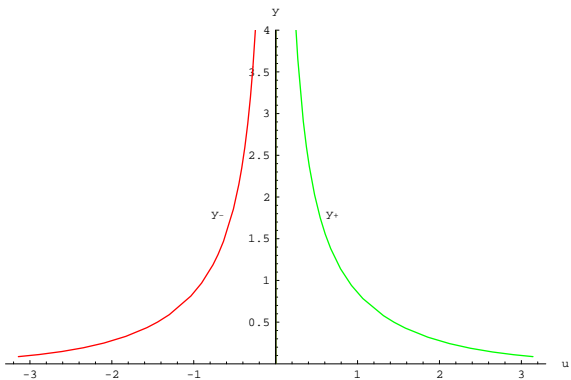}
\caption{Phantom with negative $\L$ models (case 2 of Section IV.A.1) with two branches separated by
curvature singularity (compare \cite{superjim})}
\label{fig03}
\end{figure}

Similar behaviour is present in the case 3, though now there is only one infinity
- either the scale factor collapses from infinite into a finite size in finite time, and
continues to shrink asymptotically to zero, or it expands from zero at $u=-\infty$, and
reaches infinite size in a finite time. This is shown in
Fig. \ref{fig03}. In fact, this behaviour is similar to what one
has in pre-big-bang cosmology \cite{superjim} with $y_{-}$ as a pre-big-bang branch and
$y_{+}$ as a post-big-bang branch.

The cases 4, 5 and 6 are analogous to the case 3. The appropriate formulas are obtained
from those of 2.3 in Appendix A,
but the graphs are practically like those in Fig.\ref{fig03}

\subsubsection{Phantom only models: $\L=\w=\K=0$}

\begin{figure}[h]  
\includegraphics[angle=0,scale=.46]{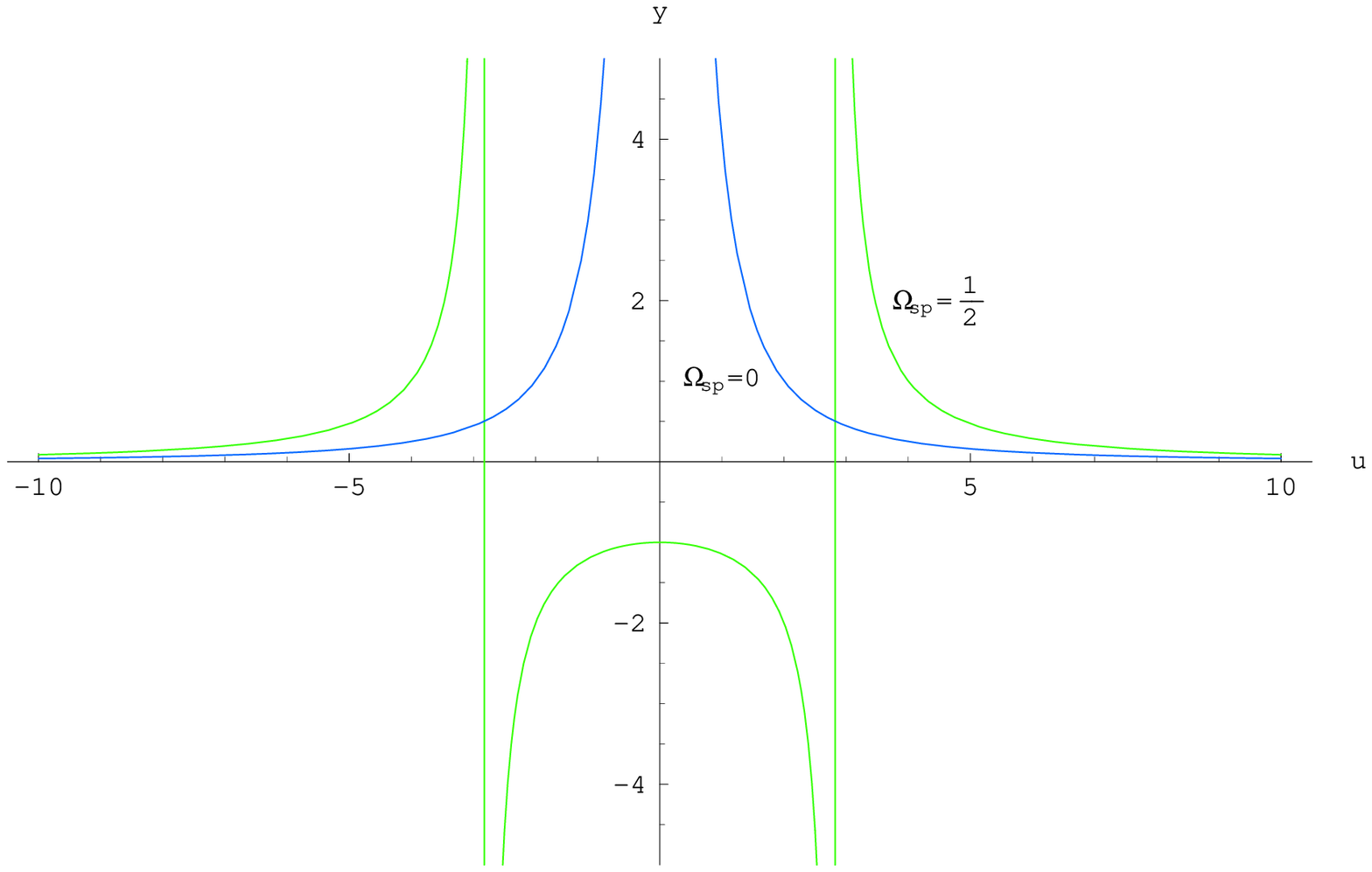}
\caption{The model (\ref{phsp}) for two phantoms only. Two cases are shown: $\sp = 0$, $\ph = 1$; $\sp = \ph = 1/2$.}
\label{fig01}
\end{figure}

If both phantoms only are present in (\ref{negpres}) we have 
\be
\label{phsp}
    y = \frac{1}{\ph \left(\frac{u-u_0}{2}\right)^2 - \frac{\sp}{\ph}}
\ee
for $\ph + \sp = 1$ and $\ph \neq 0$ (see Fig. \ref{fig01}) 
while for $\ph = 0$ we have
\be
       y_{\pm}  =  \pm \frac{1}{u-u_0},
\ee
and the behaviour of the model is as in Fig. \ref{fig03}.

\subsection{Even powers of the polynomial $Q(y)$ only models}

In this subsection we consider only the even powers of the
polynomial $Q(y)$ in (\ref{FRWbasic}), i.e.,
\begin{equation}
\label{evenonly}
\left( \frac{\ud y}{\ud u} \right)^2 = \sp y^4 + \L y^2 +
    \K + \rr y^{-2} + \st y^{-4}
\end{equation}
and make the substitution
\be
\label{zeta}
    z\equiv\frac{1}{y^2},\quad\ud\eta\equiv\frac{2\ud u}{y},
\ee
so that (\ref{evenonly}) reads as
\be
\left( \frac{\ud z}{\ud\eta} \right)^2 = \sp+\L z+\K z^2+\rr z^3+\st z^4
\ee

From now on let us assume that there is no radiation and stiff-fluid in (\ref{evenonly}), i.e., $\rr=\st=0$.
The polynomial coefficients are then (cf. Appendix A):
$a_2=\K$, $a_1=\L$, and \hbox{$a_0=\sp$}. 
The possible solutions are:
\begin{description}
\item 1'. $\K<0$, $\L>1-\sp$ (case 1.3.3).
\item 2'. \phantom{.3} $\K>0$, $\L\in\left(-2\sqrt{\sp}(1+\sqrt{\sp});2\sqrt{\sp}(1-\sqrt{\sp})
\right)$ (case 2.1).
\item 3'. $\K>0$, $\L=-2\sqrt{\sp}(1+\sqrt{\sp})$ when $\sp<1$,\\
        and $\L=-2\sqrt{\sp}(\sqrt{\sp}\pm 1)$ when $\sp>1$ (case 2.2.1).
\item 4'. $\K>0$, $\sp<1$ and $\L=2\sqrt{\sp}(1-\sqrt{\sp})$ (case 2.2.3).
\item 5'  $\K>0$, $\L<0 \land \L\notin\left[-2\sqrt{\sp}(1+\sqrt{\sp});
          2\sqrt{\sp}(1-\sqrt{\sp})\right]$ (case 2.3.1).
\item 6'  $\K>0$, $0<\L<1-\sp$ (case 2.3.5).
\end{description}

All of these contain cases where $z$ passes through zero, and that requires a closer analysis.
Firstly, because the solutions themselves change --- we are interested in
$y(u)=1/\sqrt{z(\eta(u))}$,
which, like before, may become infinite. And secondly, we are now working in a modified conformal
time $\eta$, and to analyse the solution in the cosmological time $t=u/H_0$, we need consider the
convergence of the integral
\begin{displaymath}
        u=\int\frac{\ud\eta}{2\sqrt z}.
\end{displaymath}

Fortunately, all the solutions, are of trygonometric or exponential form, so that the integral
reduces to a convergent elliptic one. In other words, the values of $u$ are finite for finite
values of $\eta$. Moreover, in some cases the integral converges with $\eta\rightarrow\infty$.
The details are given in particular cases.

Although the relation between $y$ and $z$ is different than in subsection IV.A, the qualitative properties
remain the same, that is to say, where the same classes of solutions apply.

Thus, the case 1' is again a ``bounce'' shown in Fig. \ref{fig02}.

\begin{figure}[h]
\includegraphics[angle=0,scale=.46]{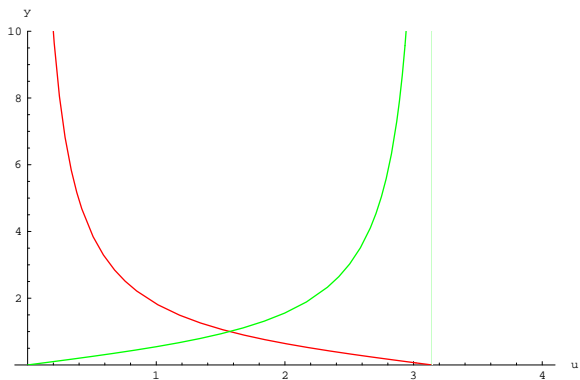}
\caption{The model with superphantom, cosmological constant and $\K \neq 0$
(solution IV.B.2').}
\label{fig05}
\end{figure}

The case 2' is a peculiar model in which the aforementioned integrals converge, and $u$ has finite values
for both the point in which the size is infinite, and when it is zero ($\eta\rightarrow\infty$).
This is depicted in Fig. \ref{fig05}.

\begin{figure}[h]
\includegraphics[angle=0,scale=.46]{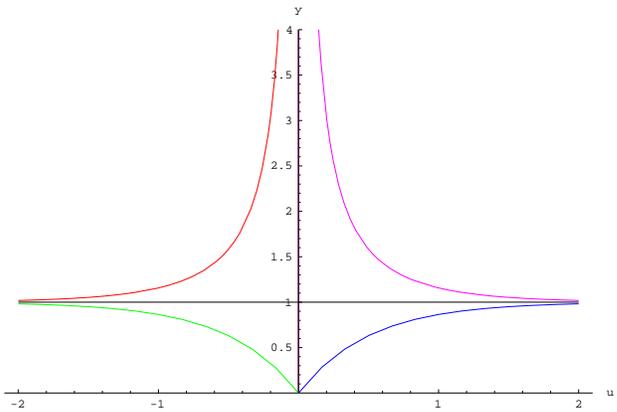}
\caption{The solution of the case B.3'. Apart from the static model (straight line in the middle) there are 
four asymptotic solutions:
two of them asymptotically approach static model in future infinity while the other two start asymptotically
with the static model at past infinity.}
\label{nowe02}   
\end{figure}

For the solutions of the case 3' the shape of the function $z(\eta)$ introduced in (\ref{zeta}) 
is given in Fig.\ref{fig221} of Appendix A with $x$ replaced by $z$ and $\eta$ replaced by 
$u$ in the graph. However, we also present these solutions in terms of the scale factor $y(u)$ in 
Fig.\ref{nowe02}. From the diagram we can see that there is a static solution $y=$ const. which corresponds to a 
double root of equation (\ref{evenonly}) with $\rr = \st = 0$. Note that the static model falls outside the classification 
given in the Appendix A since the Hubble parameter is equal to zero in this case (which is the result of the rescaled 
definition of the time parameter $u$ in (\ref{param})). Apart from the static model we have four asymptotic solutions:
two of them asymptotically approach the static model in future infinity while the other two start asymptotically 
with the static model at past infinity. 

The cases 4' and 6' are similar to the case 2' shown in Fig. \ref{fig05}.

\begin{figure}[h] 
\includegraphics[angle=0,scale=.46]{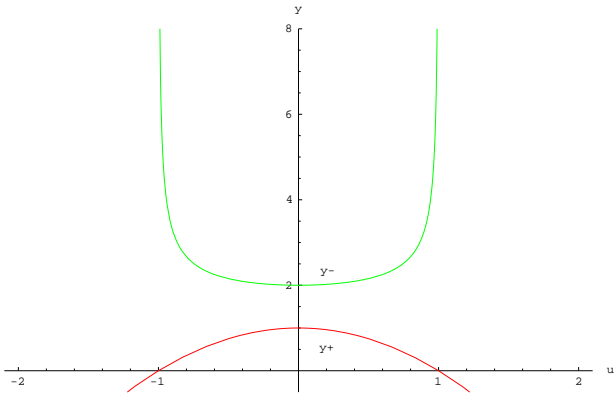}
\caption{The solutions of the case B.5'. The first is a cyclic solution from $y \propto \varrho
\to \infty$ to a minimum in $y$ and $\varrho$ and again to $y \propto \varrho
\to \infty$) and the second is a cycle in $y$ and $\varrho$ from zero to a maximum and again to zero.}
\label{nowe01}
\end{figure}

For the solutions of cases 5', the shape of the function $z(\eta)$ introduced in (\ref{zeta})
is given in Fig.\ref{fig231} of Appendix A with $x$ replaced by $z$ and $\eta$ replaced by
$u$ in the graph. In terms of the scale factor $y(u)$ these solutions are given in
Fig.\ref{nowe01}. One of them is similar to that of Fig.\ref{fig02} (a cycle from $y \propto \varrho 
\to \infty$ to a minimum in $y$ and $\varrho$ and again to $y \propto \varrho
\to \infty$) and the second describes a cycle in $y$ and $\varrho$ from zero to a maximum and again to zero.

\subsection{Phantom, walls and $\Lambda$-term models only}

This is, in fact, less general case than Case B.1, so it is a straightforward
task to apply those solutions here. We have 
\begin{equation}
\left( \frac{\ud y}{\ud u} \right)^2 = \ph y^3 + \L y^2 +
    \w y + \K + \m y^{-1},
\end{equation}
which after the change of variables  
\be
    p\equiv\frac{1}{y},\quad\ud\eta\equiv\frac{\ud u}{\sqrt{y}}\quad
\ee
gives 
\be
    \left( \frac{\ud y}{\ud\eta} \right)^2 = \ph+\L p+\w p^2+\K p^3+\m p^4 ,
\ee
and this equation is integrable in terms of elliptic functions. 

\subsection{Radiation and superphantom only models}

Leaving only the radiation pressure $\rr$ and superphantom $\sp$ non-vanishing
in (\ref{FRWbasic}) and making the substitution
\be
r\equiv y^6,\quad\ud\eta\equiv y^4\ud u ,
\ee
we get the simple equation
\be
\left(\frac{\ud r}{\ud\eta}\right)^2=\sp r+\rr ,
\ee
which solves by
\be
r=\frac{[\frac12\sp(\eta-\eta_0)]^2-1+\sp}{\sp} ,
\ee
and we have used the condition that $\rr = 1 - \sp$.

The typical evolution as considered in terms of the scale factor $y$ is similar to that 
given in Fig.\ref{fig05}. Despite the fact that the scale factor $y$ reaches either zero or 
infinity in both situations there are singularities -- for $y \to 0$ it is dominated by the 
radiation so $\varrho \to infty$ while for $y \to \infty$ it is dominated by superphantom and 
$\varrho \to \infty$, too.  

\subsection{Dust and phantom only models}

Although this case is not elementary, we present it as an important result in which we do not 
abandon the basic component of the universe. Takin all the other terms in (\ref{FRWbasic}) equal to 
zero except for dust $\Omega_{m0}$ and phantom $\Omega_{ph0}$ and making the change of the time 
coordinate $u$ as follows
\be
d\eta = \sqrt{\Omega_{ph0}} \frac{du}{\sqrt{y}}   ,
\ee
we have the equation
\be
\left(\frac{d y}{d\eta} \right)^2 = y^4 + \frac{\Omega_{m0}}{\Omega_{ph0}}  ,
\ee
which solves by 
\be
y^2 = \frac{4 \Omega_{m0} {\cal P}(\eta)}{4 \Omega_{ph0} {\cal P}^2(\eta) - \Omega_{m0}}  ,
\ee
with ${\cal P}(\eta)$ is the Weierstrass elliptic function \cite{Dabrowski96}. 

Despite this solution being elliptic, the evolution is also as in Fig.\ref{fig05}. The first 
singularity appears when $y \to 0$ and it is dominated by dust while the second singularity appears 
when $y \to \infty$ and it is dominated by phantom.

\section{Conclusion}

We have studied cosmological models with phantom $p < - \varrho$ matter which violates 
the weak energy condition. We have shown that phantom matter characterised by negative barotropic index $\gamma < 0$ allows
the curvature singularity for infinite values of the scale factor both in the past and in the future. We have also 
shown that the singularity cannot be dominated by shear so that unlike for standard $\gamma > 0$ cosmologies 
in a class of non-isotropic Bianchi type IX phantom models 
there should be no chaotic behaviour on the approach to a singularity.
We have discussed a simple model of duality of the scale factor for phantom ($\gamma < 0$) and standard 
($\gamma > 0)$ matter solutions which is alike the scale factor duality of pre-big-bang cosmology motivated by superstring 
theory duality symmetries. 

We have presented a series of exact phantom cosmologies which integrate elementary, leaving the investigation of the 
full class of models which can be integrated in terms of elliptic functions for a future paper \cite{elliptic}. 
Apart from phantom matter we have admitted most of the known types of matter content in the universe with discrete barotropic index 
$\gamma$ in the equation of state such as: dust, radiation, stiff-fluid, domain walls, cosmic strings, and 
cosmological constant. 

One interesting class of solutions we obtained contains models which start with singularity in which the scale factor 
is infinite ($a \to \infty$), then decreases, reaching some minimum $a = a_{min}$, finally expanding again into another 
singularity where ($a \to \infty$). In the standard picture for $\gamma > 0$ matter, the universe follows the
sequence: expansion -- maximum -- recollapse. On the other hand, for phantom $\gamma < 0$ models 
one has the opposite sequence: collapse -- minimum -- expansion. 
Another interesting class of solutions is characterised by either an expansion from zero energy density to a future curvature 
singularity, or a collapse from the infinite value of the scale factor into a state of zero energy density. There exists also static solutions 
(generalized Einstein Static Universes) together with asymptotic solutions though asymptotic solutions approach static model either from 
singularity of the energy density (in which the scale factor $a \to \infty$), or from the state of zero energy density (in which the scale factor 
$a \to 0$).

Finally, in the models which contain a mixture of phantom and standard matter (e.g. dust, radiation) there exist two types of singularities 
(with the energy density $\varrho \to \infty$): one is dominated by phantom with the scale factor $y \to \infty$ and another 
is dominated by standard matter with the scale factor $y \to 0$.

We remark that phantom cosmologies form an interesting set of the models of the universe which do not contradict 
observational data of supernovae type Ia and that their mathematical and physical properties deserve further studies. 

\section{Acknowledgements}

The support from Polish Research Committee (KBN) grants No 2P03B 090 23 (M.P.D.)
and No 2P03B 107 22 (M.S.) is acknowledged.

\appendix
\section{General classification of solutions to the canonical equation}

We take into account the equation
\begin{equation}
        \left(\frac{\ud x}{\ud u}\right)^2 = a_2x^2+a_1x+a_0 = W(x) \label{basic}
\end{equation}
together with the constraint:
\begin{equation}
        a_2+a_1+a_0=1, \label{cond}
\end{equation}
and seek its solutions, that might depict the evolution of the Universe, that is, such that $x \geq 0$.
We also define $e_1$ and $e_2$ to be the roots of the quation $W(x)=0$, and $\Delta=a_1^2-4a_0a_2$.
The main constraint on $y$ is the nonnegativity of the polynomial $W(x)$ implied by equation
(\ref{basic}).

The general solution of the equation (\ref{basic}) is
\begin{equation}
        \frac{2a_2 x+a_1}{\sqrt{\Delta}} = \cosh[\sqrt{a_2}(u-u_0)]
\end{equation}
where $u_0$ can, in general, be complex. The graphs depicting the particular cases, were drawn with
$u_0=0$, except for case 2.3.1 whose solution requires $u_0= i\pi/2\sqrt{a_2}$. It should be
kept in mind, that the solutions may be arbitrarily shifted in the direction of $u$-axis.

\subsection{$a_2<0$}
\subsubsection*{{\rm 1.1.} $\Delta<0$}
    This makes evolution impossible, as $W(x)<0$ for all real values of $y$. Writing
$\Delta=a_1^2+4a_0a_1+4a_0(a_0-1)$, we get its discriminant $\widetilde{\Delta}=16a_0$, and the roots:
$a_{1\mp}=-2\sqrt{a_0}(\sqrt{a_0}\pm 1)$. Clearly, $a_0 < 0$, would leads to $\Delta>0$, and $a_0=0$ implies
$a_2=0$ which contradicts (\ref{cond}). For $a_0>0$, we would need $a_1\in(a_{1-};a_{1+})$, but this
does not hold together with (\ref{cond}).

Such case is impossible with the assumptions made.

\subsubsection*{{\rm 1.2.} $\Delta=0$}
        As above we could have $\widetilde\Delta=0$, but this would lead to $a_0=a_1=0$, and break (\ref{cond}).
If $\widetilde\Delta>0$ the only possibilities, namely $a_1=a_{1\mp}$, are ruled out by the same condition.

\subsubsection*{{\rm 1.3.} $\Delta>0$}
        Here we could have $a_0>0$, and $a_1>a_{1+}$ or $a_1<a_{1-}$ which, reasoning as above, gives
in the end $a_1=1-a_2-a_0>1-a_0$. Alternatively, $a_0\leq 0$, leads directly to $a_1>1-a_0$.

\subsubsection*{{\rm 1.3.1.} $e_1>e_2>0$}
        This requires $a_1>0,a_0<0$ which can hold with together with $a_1>1-a_0$. The solution is periodic
in the form:
\begin{equation}
        x = \frac{\sqrt{\Delta}\cos[\sqrt{|a_2|}(u-u_0)]-a_1}{2a_2} \label{period}
\end{equation}

\subsubsection*{{\rm 1.3.2.} $e_1>e_2=0$}

Necessarily $a_0=0$ and $a_1=1-a_2>1$. The solution is simply:
\begin{equation}
        y = \frac{a_1}{2a_2}\left(\cos[\sqrt{|a_2|}(u-u_0)]-1\right)
\end{equation}
It passes through zero, unlike the previous one.

\subsubsection*{{\rm 1.3.3.} $e_1>0>e_2$}

The conditions here are: $a_0>0$ and $a_1 > 1-a_0$. The solution now passes through zero twice,
which means it is no longer periodic, but depicts a single ``cycle''. The formula is identical to (\ref{period}).

\subsubsection*{{\rm 1.3.4.} $0\geq e_1>e_2$}

\begin{figure}[h]
\includegraphics[angle=0,scale=.46]{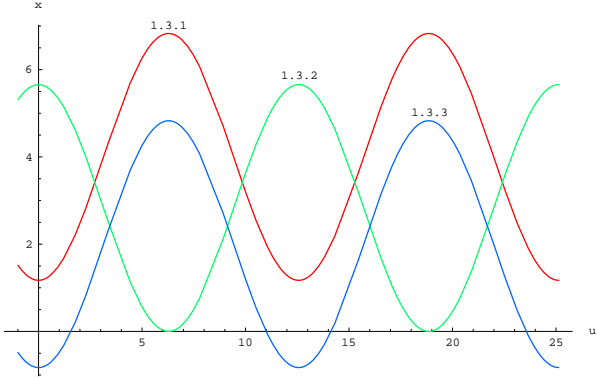}
\caption{Cases 1.3.1, 1.3.2 and 1.3.3}
\label{fig131}
\end{figure}

This case would require $a_0\leq 0$ and $a_1<0$, which cannot hold together with (\ref{cond}). The plots 
of 1.3.1, 1.3.2 and 1.3.3 are given in Fig.\ref{fig131}.

\subsection{$a_2>0$}

\subsubsection*{{\rm 2.1.} $\Delta<0$}

\begin{figure}[h]
\includegraphics[angle=0,scale=.46]{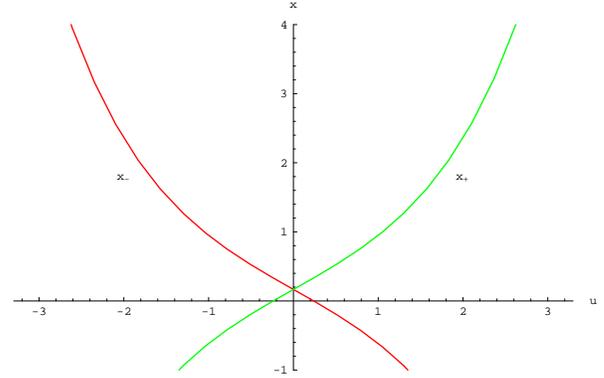}
\caption{Case 2.1 }
\label{fig21}
\end{figure}

This is only possible with $\widetilde\Delta>0 \Rightarrow a_0>0,\;a_1\in(a_{1-};a_{1+})$, which
is now a stronger restriction than $a_1=1-a_2-a_0<1-a_0$. There are two solutions now, depending on the
choice of the derivative sign in equation (\ref{basic}):
\begin{equation}
        x_{\pm} = \frac{\pm\sqrt{|\Delta|}\sinh[\sqrt{a_2}(u-u_0)]-a_1}{2a_2}
\end{equation}
These are respectively, monotonic expansion (from zero size to infinity in infinite time)
or monotonic collapse (in finite time from finite size). This is plotted in Fig.\ref{fig21}.

\subsubsection*{{\rm 2.2.} $\Delta=0$}
        Note, that $a_0<0$ would lead to $\widetilde\Delta<0$ and hence $\Delta>0$, and this is not the case.

\subsubsection*{{\rm 2.2.1.} $e_1=e_2>0$}

        $a_0>0$, so $a_1=a_{1-}<0$, and for $a_0>1$ we also have \hbox{$a_1=a_{1+}<0$}.
        For \hbox{$x>e_1=\frac{-a_1}{2a_2}$} the evolution is either an unbounded
expansion, or an asymptotical collapse.
\begin{equation}
        x_{\pm}=\frac{-a_1}{2a_2}(1+e^{\pm\sqrt{a_2}(u-u_0)})
\end{equation}

\begin{figure}[h]
\includegraphics[angle=0,scale=0.46]{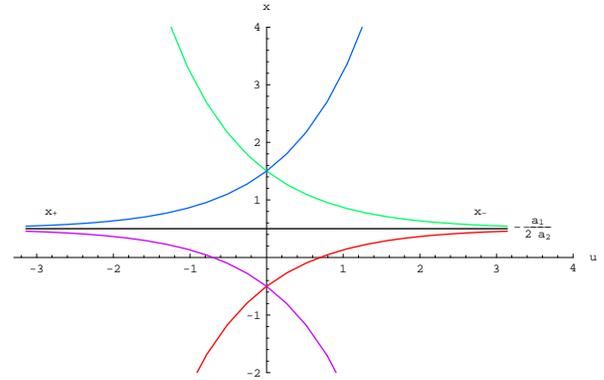}
\caption{Case 2.2.1. Case 2.2.2 may be pictured if $a_1=0$; case 2.2.3
        when $- a_1/2a_2 <0$.}
\label{fig221}
\end{figure}

When $0\leq x<e_1$, the expansion is asymptotical, and collapse occurs in finite time.
\begin{equation}
        x_{\pm}=\frac{-a_1}{2a_2}(1-e^{\pm\sqrt{a_2}(u-u_0)})
\end{equation}
Also, $x=e_1$ is a static solution which is unstable due to nearby asymptotic solutions.
The plot of 2.2.1 is given in Fig.\ref{fig221}.

\subsubsection*{{\rm 2.2.2.} $e_1=e_2=0$}

We immediately get $a_1=a_0=0$ which yields the following solution:
\begin{equation}
        x_{\pm}=e^{\pm\sqrt{a_2}(u-u_0)}.
\end{equation}

\subsubsection*{{\rm 2.2.3.} $e_1=e_2<0$}

$a_1=a_{1+}>0$, requiring $a_0<1$.
This case is almost identical to 2.2.1, only the root is now negative, eliminating the
solutions $x \leq e_1$.
The only possibility for collapse is to happen in finite time now.

\subsubsection*{{\rm 2.3.} $\Delta>0$}
        The condition $a_1<1-a_0$ applies to all the subcases here. Depending on $a_0$ there are also further restrictions:
\begin{description}
\item{$a_0>0\;\Rightarrow$} $\widetilde\Delta>0$, so $a_1\notin[a_{1-};a_{1+}]$.
\item{$a_0=0\;\Rightarrow$} $a_1 \ne 0$.
\item{$a_0<0\;\Rightarrow$} No other conditions.
\end{description}
        In general, the situation resembles that in 1.3. All the graphs look the same the only difference being a shift
in the $y$ direction.

\subsubsection*{{\rm 2.3.1.} $e_1>e_2>0$}

\begin{figure}[h]
\includegraphics[angle=0,scale=.46]{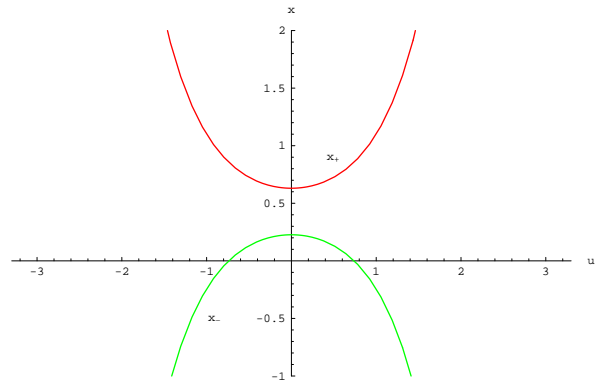}
\caption{Case 2.3.1. Other cases 2.3.2, 2.3.3, 2.3.4, and 2.3.5 are obtained when the graph is
        shifted down in the direction $x$ so that $x_-$ passes through zero, $x_-<0$ and $x_+>0$,
        $x_+$ passes through zero, and $x_+$ intersects the abscissa at two points,
        respectively.}
\label{fig231}
\end{figure}

$a_0>0$, and $0>a_1$ give two admissible regions of evolution. For $x \geq e_1$, we have a ``bounce'' (infinite
size is reached in infinite time, though). When $0 \leq x \leq e_1$, the evolution is at most one ``cycle'' similar
to that of 1.3.3. To be exact:
\begin{equation}
        x_{\pm}=\frac{\pm\sqrt\Delta\cosh[\sqrt{a_2}(u-u_0)]-a_1}{2a_2}.
\end{equation}
The plot of 2.3.1 is given in Fig.\ref{fig231}.

\subsubsection*{{\rm 2.3.2.}  $e_1>e_2=0$}

This is almost identical to the previous case, only now $x_-$ passes through zero,
which is its only nonnegative value,
thus giving rise to a static universe of zero size. $x_+$ behaves as above.
Here, we have: $a_0=0$ and $a_1<0$.

\subsubsection*{{\rm 2.3.3.} $e_1>0>e_2$}

$a_0<0$ and $a_1<1-a_0$ now. Again, it is like 2.3.1 with $x_-$ discarded.

\subsubsection*{{\rm 2.3.4} $e_1=0>e_2$}

$a_0=0$ and $0<a_1<1$. This is another ``bounce'' solution which passes through
a possible singularity at $x=0$.

\subsubsection*{{\rm 2.3.5.}  $0>e_1>e_2$}

$a_0>0$ and $0<a_1<1-a_0$. The solution is separated into two independent ones,
describing expansion or collapse, both going through zero.

\end{document}